\documentclass[nofootinbib,showpacs,preprintnumbers,amsmath,amssymb,floatfix]
{revtex4}
\usepackage{epsf}

\begin{document}


\title{Perturbative  $Q^2$-power corrections to the structure function $g_1$.}

\vspace*{0.3 cm}

\author{B.I.~Ermolaev}
\affiliation{Ioffe Physico-Technical Institute, 194021
 St.Petersburg, Russia}
\author{M.~Greco}
\affiliation{Department of Physics and INFN, University Rome III,
Rome, Italy}
\author{S.I.~Troyan}
\affiliation{St.Petersburg Institute of Nuclear Physics, 188300
Gatchina, Russia}

\begin{abstract}
 We show that  ($\sim 1/(Q^2)^k$) power corrections to the
spin structure function $g_1$ at small $x$ and large $Q^2$ are
generated perturbatively from the regulated infrared divergencies.
When $Q^2$ are small, the corrections are $\sim (Q^2)^k$. We
present the explicit series of such terms as well as the formulae
for their resummation. These contributions are not included in the
standard analysis of the experimental data. We argue that
accounting for such terms can sizably change the impact of the
power corrections conventionally attributed to the higher twists.
\end{abstract}


\maketitle

\section{Introduction}
The theoretical description of the $Q^2$ -dependence of the structure
function $g_1$ in  perturbative QCD  is mostly performed
for the kinematic region of large $Q^2$. However, for the phenomenological
analysis  of the results of the COMPASS collaboration  one needs
the possible knowledge of $g_1$ at small $x$ and small $Q^2$ (see e.g.
Ref.~\cite{compass}). Let us remind that the conventional Standard
Approach (SA) based on combining the DGLAP evolution
equations\cite{dglap} with  phenomenological inputs for the initial
parton densities\cite{fits}, cannot be used for description of
$g_1$ in this region. Strictly speaking, the SA can be applied only
for large values of $x$ ($x\sim 1$) and large $Q^2$: $Q^2 \gg
Q^2_0$, where $Q^2_0$ is the starting point of the $Q^2$
-evolution.
Indeed the small-$x$ region
lies beyond the reach of the SA because DGLAP does not include the
total resummation of $\propto\ln^k 1/x$ terms. In order to
describe the experimental data at small $x$, in the SA one has to include
singular $\propto x^{-\alpha}$ terms in the expressions for the initial
parton densities. Such factors act as the leading singularities
(simple poles) in the Mellin transform of $g_1(x)$ and provide
$g_1$ with the Regge asymptotics: $g_1\sim x^{-\alpha}$ when $x\to
0$.

On the other hand, in our approach\cite{egtns, egts} based on the total resummation
of the leading $\ln^k 1/x$ terms in perturbative
QCD, the Regge behaviour of $g_1$ at $x \to 0$  appears naturally,
 independent of the value of $Q^2$. Indeed it results
from the leading singularities of the  anomalous dimensions and
coefficient functions, which are  branching points and
not simple poles. The singularities of the anomalous
dimensions and coefficient functions coincide. This is very
important because it guarantees the  independence  of $Q^2$
of the intercepts of $g_1$. Furthermore by fitting the experimental data
in our approach one does  not need  ad hoc singular factors $\sim
x^{-\alpha}$ in the  initial parton densities.

The theoretical aspects of the power corrections to the DIS
structure functions were considered in Refs.~\cite{pc}
-\cite{ster} at large $x$ and $Q^2$. The interplay between the
perturbative and non-perturbative corrections in the hard
kinematics was recently considered in detail in Ref~\cite{stef}.
In fitting experimental data in the small-$x$ region, the
discrepancies from the SA predictions are conventionally
interpreted (see e.g. Refs.~\cite{sid}) as the (non-perturbative)
higher twist power ($\sim (1/Q^2)^k$) contributions. However as
the small-$x$ region is beyond the reach of SA, the real size of
the higher twist corrections can be erroneously overestimated. It
is clear that systematic account of power corrections
$\sim(1/Q^2)^k$ can be satisfactory only when using formulas  that
already account for the resummation of logarithmic contributions.

We have recently proposed in Ref.~\cite{egtsmq} a generalization
of our previous results\cite{egtns,egts} for $g_1$. Although this
extension goes beyond the logarithmic accuracy we keep, it looks
quite natural. In addition to the standard region of large $Q^2$,
this generalization can describe the small-$Q^2$ region, though in
a model-dependent way. Our suggestion, based on the analysis of
the Feynman graphs for $g_1$ at small $x$, is  to replace $Q^2$ by
$Q^2 + \mu^2$  in our previous formulas, with $\mu$ being the
infrared cut-off
Such a shift also leads  to replacement the variable $x$ by$x' = x
+ \mu^2/2pq$. The variable $x'$ is similar to the Nachtmann
variable\footnote{We are grateful to L.N.~Lipatov for reminding
this.}.

In the present paper we show that regulating the infrared
divergencies originates also the power corrections for $g_1$ at
small $x$. They differ from the well-known power
corrections\cite{pc,korst,ster} related to the resummation of the
Sudakov logarithms: first, they come from the ladder Feynman
graphs in the Regge kinematics where virtual gluons are not always
soft; then, in contrast to Refs.~\cite{pc,korst,ster,stef}, we
never use\footnote{It is known (see Ref.~\cite{egtalpha}) that
this parameterizations fails for $g_1$ at small $x$} the
parametrization $\alpha_s = \alpha(Q^2)$ and do not consider
inclusions of power terms\cite{shirk} into the standard
expressions for $\alpha_s$. We present the explicit formulas for
the total resummation of these new corrections  and show that at
large $Q^2$ they are $\sim 1/(Q^2)^k$ but in the small- $Q^2$
region they are $\sim (Q^2)^k$.
The paper is organized as follows: in Sect.~2 we explain why there
should be the infrared perturbative and non-perturbative power
corrections to $g_1$. We also discuss in this Sect. the difference
in the infrared properties of $g_1$ at large and small $x$. In
Sect.~3 we remind the essence of the model for $g_1$ suggested in
Ref.~\cite{egtsmq} and list explicit expressions for the singlet
and non-singlet $g_1$ at the kinematic region of small $x$ and
arbitrary $Q^2$. In Sect.~4 we give in more detail than in
Ref.~\cite{egtsmq} theoretical arguments in favor of those
expressions. The expressions for $g_1$ enlisted in Sect.~3 and
proved in Sect.~4 account for the total resummation of leading
logarithms of $x$ and $Q^2$ when $Q^2$ are large but they are also
valid for small $Q^2$. They implicitly include the power $Q^2$
-corrections. The explicit expressions for the power corrections
are extracted from those formulas in Sect.~5. We show here that
the power corrections at large and small $Q^2$ are quite different
but the lowest twist contribution to $g_1$ is always leading
regardless of $Q^2$. Finally, Sect.~6 is for discussion.

\section{Origin of the infrared-dependent contributions to $g_1$}

It is well-known that the DIS structure functions are introduced
through  the hadronic tensor $W_{\mu\nu}$. In particular, the
spin-dependent part of $W_{\mu\nu}$ for the electron-proton DIS is
parameterized by the structure functions $g_1$ and $g_2$:
\begin{equation}\label{wpol}
W_{\mu\nu}^{spin} = \imath \epsilon_{\mu\nu\lambda\rho} \frac{M
q_{\lambda}}{pq} \Big[ S_{\rho}~g_1(x, Q^2) + \Big(S_{\rho} -
\frac{Sq}{pq}\Big)~g_2(x, Q^2)\Big]~
\end{equation}
where $p,M,S$ are the proton momentum, mass and spin respectively,
$q$ is the virtual photon momentum. The spin structure functions
$g_{1,2}$ have the non-singlet, $g_{1,2}^{NS}$, and singlet,
$g_{1,2}^S$, components. In general, $g_{1,2}$ (and all other DIS
structure functions) acquire both perturbative and
non-perturbative QCD contributions.

In the first place, there is a non-perturbative term
$W_{\mu\nu}^{NPT}$ in the region of small $Q^2$. For instance,
there are known examples of the lattice calculations for some
structure functions (see e.g. Ref.~\cite{mart}), although
basically $W_{\mu\nu}^{NPT}$ is poorly known.

The other part, $W^{PT}_{\mu\nu}$, of $W_{\mu\nu}$ includes both
perturbative and non-perturbative QCD contributions. The standard
way to calculate the structure functions is using the
factorization. According to it, $W_{\mu\nu}$ is regarded as a
convolution
\begin{equation}\label{conv}
W^{PT}_{\mu\nu} =W_{\mu\nu}^q \otimes \Phi_q + W_{\mu\nu}^g
\otimes \Phi_g
\end{equation}
of the partonic tensors $W_{\mu\nu}^q,~W_{\mu\nu}^g $ (where $q,g$
label the incoming parton, i.e. a quark or a gluon respectively)
and probabilities $\Phi_{q,g}$ to find this parton (quark or
gluon) in the incoming hadron. The probabilities $\Phi_{q,g}$
include both perturbative and non-perturbative contributions
whereas purely perturbative tensors $W_{\mu\nu}^{q,g}$ describe
their $x$- and $Q^2$- evolutions. There are no known explicit
expressions for $\Phi_{q,g}$. Instead, they are approximated by
the initial parton densities $\delta q$ and $\delta g$ defined
from fitting experimental data at $Q^2  \sim 1$~GeV$^2$ and $x
\sim 1$. The partonic tensors $W_{\mu\nu}^q,~W_{\mu\nu}^g $ evolve
these densities into the region where $Q^2 \gg \mu^2$ and $x \ll
1$, with $\mu^2$ being the starting point of the $Q^2$ -evolution.
Such a contribution to $W_{\mu\nu}$ is called the lowest twist
contribution $W^{LT}_{\mu\nu}$. Besides, there are the higher
twists contributions $W^{HT}_{\mu\nu}$. They can be interpreted
either in terms of more involved convolution or as essentially
non-perturbative objects.

In order to calculate $W^{q,g}_{\mu\nu}$, one has to regulate
infrared (IR) singularities in involved Feynman graphs. Such a
regulation is different for large and small $x$. At large $x$ one
can use DGLAP. In DGLAP this problem is solved with assuming a
non-zero virtuality $\mu^2$ for the initial partons and imposing
the ordering
\begin{equation}\label{order}
\mu^2 < k^2_{1~\perp} < k^2_{2~\perp}... < k^2_{n~\perp} < Q^2
\end{equation}
on the transverse momenta of the ladder partons (the numeration in
Eq.~(\ref{order}) runs from the bottom to the top of the ladders).
Eq.~(\ref{order}) manifests that $k_{r~\perp}$ acts as an infrared
cut-off for integrating over $k_{r+1}$. However, the ordering
allows to collect the contributions that are leading at large $x$
only. In order to account for the leading (double-logarithmic)
contributions at $x \ll 1$, the upper limit of the integration in
Eq.~(\ref{order}) should be changed for $ (p + q)^2 \approx 2pq$
and the ordering should be lifted. Without the ordering, the IR
singularities in the ladder rungs are not any longer regulated
automatically, so an infrared cut-off should be introduced for
integration over every loop momentum. It is the reason why the
$\mu$ -dependence of $g_1$ is getting more involved at small $x$.
Usually the cut-off is identified with $\mu$, the starting point
of the $Q^2$ -evolution, though it is not mandatory. Obviously,
the value of $\mu$ should be large enough to justify using the
Perturbative QCD:

\begin{equation}
\label{mu} k^2_i > \mu^2 > \Lambda^2_{QCD}.
\end{equation}
Generally speaking, there are different ways to introduce infrared
cut-offs but providing IR-divergent propagators with fictiones
masses is most wide-spread. Obviously, no observables should
depend on a value of $\mu$ and ways of its introducing. It means
that the explicit $\mu$ -dependence in $W_{\mu\nu}^{q,g}$ should
be compensated by a $\mu$ -dependence of $\delta q,~\delta g$.
However, the latter are known as phenomenological expressions
containing a set of numerical parameters fixed from fitting
experimental data. Those parameters are supposed to be $\mu$
-dependent, though in unexplicit way.

Before considering the IR properties of $g_1$ at small $x$, let us
discuss the well-known expression for the non-singlet $g_1$ in the
Standard Approach:
\begin{equation}\label{mellinsa}
g_1^{NS}(x, Q^2, \mu^{2})= (e^2_q/2) \int_{- \imath
\infty}^{\imath \infty} \frac{d \omega}{2 \pi \imath} x^{- \omega}
C_q(\omega) \delta q (\omega) \exp \Big[ \int_{\mu^2}^{Q^2}
\frac{d k^2_{\perp}}{k^2_{\perp}} \gamma_{qq} \big(k^2_{\perp},
\alpha_s(k^2_{\perp})\big)\Big]
\end{equation}
where $e_q$ is the quark electric charge, $C_q$ is the coefficient
function and $\gamma_{qq}$ is the DGLAP non-singlet anomalous
dimension. When $\gamma_{qq}$ is taken in LO, the integral in
Eq.~(\ref{mellinsa}) is known to be $(1/b) \ln [\ln
(Q^2/\mu^2)/\ln(\mu^2/\Lambda^2_{QCD})]$, with $b$ being the first
coefficient of the Gell-Mann -Low function. The expression for the
singlet $g_1$ looks similar, though  more involved. Both the
coefficient functions and the anomalous dimensions in SA are known
in few first orders in $\alpha_s$. It means that using
Eq.~(\ref{mellinsa}) is theoretically based for the kinematic
region where $x$ are not far from 1 and $Q^2 \gg \mu^2 \approx
1$~GeV$^2$. The expression for $g_1$ in Eq.~(\ref{mellinsa})
depends on the value of $\mu$. This dependence is supposed to
disappear when $g_1$ is complemented by a contribution $g_1^{HT}$
extracted from $W_{\mu\nu}^{HT}$.  In practice, the treating
experimental data on Polarized DIS is carried out as follows (see
e.g. the recent paper Ref.~\cite{sid} and refs therein): the data
are compared with $g_1$ of Eq.~(\ref{mellinsa}) and the
discrepancy is attributed to the higher twists impact. However, SA
for $g_1$ is reliable at
 large $x$. At the small-$x$ region it should be modified. In the
 next section we present explicit expressions for $g_1^S$ and
 $g_1^{NS}$ replacing the DGLAP expressions in the small-$x$ region.

\section{Expressions for $g_1$ at small $x$ and arbitrary $Q^2$}

When $x \ll 1$, the contributions $\sim \ln^k(1/x)$ are large, so
they should be accounted for to all orders in the QCD coupling.
The total resummation of the leading logarithms of $x$ for $g_1$
was done in Refs.~\cite{egts,egtns} for the region of $Q^2 \gtrsim
\mu^2$. We remind that, contrary to DGLAP, the expressions for
$g_1$ in Refs.~\cite{egts,egtns} are valid both for large $Q^2$
and for $Q^2 \sim \mu^2$. Recently, in Ref.~\cite{egtsmq} we have
suggested a simple prescription to generalize those results to
arbitrary values of $Q^2$: in the formulas of
Refs.~\cite{egts,egtns} $Q^2$ should be replaced by $Q^2 + \mu^2$.
This conclusion follows from the observation that the
contributions of Feynman graphs to $g_1$ at small $x$ depend on
$Q^2$ through $Q^2 + \mu^2$ only. It automatically leads to the
shift $x \to x + z$, with $z = \mu^2/2pq$. As the prescription is
beyond the logarithmic accuracy that we kept in our previous
papers, we call it a model. The theoretical grounds of this model
are given in the next Sect. According to our results, the
non-singlet, $g_1^{NS}$, component of $g_1$ at the small-$x$
region is given by the following expression:

\begin{equation}
\label{gnsint} g_1^{NS}(x+z, Q^2+\mu^2) = (e^2_q/2) \int_{-\imath
\infty}^{\imath \infty} \frac{d \omega}{2\pi\imath }\Big(
\frac{1}{z + x} \Big)^{\omega} C_{NS}(\omega) \delta q(\omega)
\Big(\frac{Q^2 + \mu^2}{\mu^2} \Big)^{H_{NS}(\omega)}~,
\end{equation}
with the new coefficient functions  $C_{NS}$,
\begin{equation}
\label{cns} C_{NS}(\omega) =\frac{\omega}{\omega - H_{NS}(\omega)}
\end{equation}
and the anomalous dimensions $H_{NS}$,
\begin{equation}
\label{hns} H_{NS} = (1/2) \Big[\omega - \sqrt{\omega^2 -
B(\omega)} \Big]
\end{equation}
where
\begin{equation}
\label{b} B(\omega) = (4\pi C_F (1 +  \omega/2) A(\omega) +
D(\omega))/ (2 \pi^2)~.
\end{equation}
 $ D(\omega)$ and $A(\omega)$ in Eq.~(\ref{b}) are
expressed in terms of  $\rho = \ln(1/x)$, $\eta =
\ln(\mu^2/\Lambda^2_{QCD})$, $b = (33 - 2n_f)/12\pi$ and the color
factors
 $C_F = 4/3$, $N = 3$:

\begin{equation}
\label{d} D(\omega) = \frac{2C_F}{b^2 N} \int_0^{\infty} d \rho
e^{-\omega \rho} \ln \big( \frac{\rho + \eta}{\eta}\big) \Big[
\frac{\rho + \eta}{(\rho + \eta)^2 + \pi^2} + \frac{1}{\rho +
\eta}\Big] ~,
\end{equation}

\begin{equation}
\label{a} A(\omega) = \frac{1}{b} \Big[\frac{\eta}{\eta^2 + \pi^2}
- \int_0^{\infty} \frac{d \rho e^{-\omega \rho}}{(\rho + \eta)^2 +
\pi^2} \Big].
\end{equation}
$H_{NS}$  and $C_{NS}$ account for DL and SL contributions to all
orders in $\alpha_s$ and, contrary to the DGLAP phenomenology,
$\delta q$ does not contain singular factors. The infrared cut-off
$\mu$ obeys Eq.~(\ref{mu}). Expression (\ref{gnsint}) is valid for
large $Q^2$, i.e. for $Q^2 \gg \mu^2$ where $x \gg z$ and for
small $Q^2,~Q^2 \leq \mu^2$ where $x \leq z$. The expression for
the singlet component, $g_1^S$, of $g_1$ is more involved:
\begin{equation}\label{g1int}
g_1^S = g_1^{(+)} + g_1^{(-)},
\end{equation}
with

\begin{equation}\label{g1pm}
g_1^{(\pm)} = \frac{<e^2_q>}{2} \int_{- \imath \infty}^{\imath
\infty} \frac{d \omega}{2 \pi \imath} \Big(\frac{1}{z + x}
\Big)^{\omega}
\Big(C_q^{(\pm)} \delta q -
\frac{A'}{2 \pi \omega^2} C_g^{(\pm)} \delta g \Big)
\Big(\frac{Q^2 + \mu^2}{\mu^2} \Big)^{\Omega_{(\pm)}}
\end{equation}
where $<e^2_q>$ stands for the sum of electric charges: $<e^2_q> =
10/9$ for $n_f = 4$, $\delta q$ and
$\delta g$ are the initial quark and gluon densities.

The exponents $\Omega_{(\pm)}$ and coefficient functions
$C^{(\pm)}_{q,g}$ are:
\begin{equation}
\label{omegapm}
\Omega_{(\pm)} = \frac{1}{2} \big[ H_{qq} +  H_{gg} \pm R \big]~.
\end{equation}
\begin{eqnarray}\label{cbk}
C_q^{(+)} &=& \frac{\omega}{RT}
\Big[
(H_{qq}-\Omega_{(-)})(\omega-H_{gg}) + H_{qg}H_{gq} +
H_{gq}(\omega-\Omega_{(-)})
\Big]~, \\ \nonumber
C_q^{(-)} &=& \frac{\omega}{RT}
\Big[
(\Omega_{(+)}-H_{qq})(\omega-H_{gg}) - H_{qg}H_{gq} +
H_{gq}(\Omega_{(+)}-\omega)
\Big]~, \\ \nonumber
C_g^{(+)} &=& \frac{\omega}{RT}
\Big[
(H_{gg}-\Omega_{(-)})(\omega-H_{qq}) + H_{qg}H_{gq} +
H_{qg}(\omega-\Omega_{(-)})
\Big] \Big(-\frac{A'}{2\pi\omega^2}\Big)~, \\ \nonumber
C_g^{(-)} &=& \frac{\omega}{RT}
\Big[
(\Omega_{(+)}-H_{gg})(\omega-H_{qq}) - H_{qg}H_{gq} +
H_{qg}(\Omega_{(+)}-\omega)
\Big] \Big(-\frac{A'}{2\pi\omega^2}\Big)~.
\end{eqnarray}
Here
\begin{equation}\label{rt}
R= \sqrt{ (H_{qq} -  H_{gg})^2 + 4 H_{qg} H_{gq}}~,\qquad
T= \omega^2 - \omega (H_{gg} + H_{qq}) + (H_{gg}H_{qq} - H_{gq}H_{qg})~
\end{equation}
and
\begin{eqnarray}\label{hik}
&& H_{qq} = \frac{1}{2} \Big[ \omega + Z + \frac{b_{qq} -
b_{gg}}{Z}\Big],\qquad H_{qg} = \frac{b_{qg}}{Z}~, \\ \nonumber
&& H_{gg} = \frac{1}{2} \Big[ \omega + Z - \frac{b_{qq} -
b_{gg}}{Z}\Big],\qquad H_{gq} =\frac{b_{gq}}{Z}~
\end{eqnarray}
where
\begin{equation}
\label{z}
 Z = \frac{1}{\sqrt{2}}\sqrt{(\omega^2 - 2(b_{qq} + b_{gg})) +
\sqrt{(\omega^2 - 2(b_{qq} + b_{gg}))^2 - 4 (b_{qq} - b_{gg})^2 -
16b_{gq} b_{qg} }}
\end{equation}
with
\begin{equation}\label{bik}
b_{ik} = a_{ik} + V_{ik}~.
\end{equation}
The Born contributions $a_{ik}$ are defined as follows:
\begin{equation}\label{aik}
a_{qq} = \frac{A(\omega) C_F}{2 \pi}~,\quad a_{gq} = -\frac{n_f
A'(\omega)}{2 \pi}~,\quad a_{qg} = \frac{ A'(\omega)
C_F}{\pi}~,\quad a_{gg} = \frac{4 N A(\omega)}{2 \pi}~.
\end{equation}
At last, non-ladder contributions are:
\begin{equation}\label{vik}
V_{ik} =  \frac{m_{ik}}{\pi^2} D(\omega)~,
\end{equation}
with
\begin{equation}\label{mik}
m_{qq} = \frac{C_F}{2 N}~,\quad m_{gg} = - 2N^2~,\quad
m_{gq} = n_f \frac{N}{2}~,\quad m_{qg} = - N C_F~.
\end{equation}

The additional factor $\Big(-\frac{A'}{2\pi\omega^2}\Big)$ in the
coefficients $C_g^{(\pm)}$ in Eqs.(\ref{cbk}),  with
\begin{equation}\label{aprime}
A'(\omega) = \frac{1}{b} \Big[ \frac{1}{\eta} - \int_{0}^{\infty}
\rho \frac{d \rho e^{- \omega \rho}}{(\rho + \eta)^2} \Big]~,
\end{equation}
is the small-$\omega$ estimate for the quark box diagram that
dominates in the Born term relating initial gluons to the
electromagnetic current. $A'(\omega)$ is the Mellin representation
of the QCD running coupling $\alpha_s$ involved in the quark box.

Besides resummation of the leading logarithms,
Eqs.~(\ref{gnsint},\ref{g1int}) differ from DGLAP in the
parametrization of $\alpha_s$: the DGLAP -prescription is
$\alpha_s = \alpha_s(Q^2)$ whereas in our approach $\alpha_s$ is
replaced by $A$ and $A'$ defined in Eqs.~(\ref{a},\ref{aprime}).
Such a difference results into a drastic difference in the form of
the $Q^2$ -dependence of $g_1$ between our approach and SA:
instead of the factor $(Q^2+ \mu^2)/\mu^2$ in
Eqs.~(\ref{gnsint},\ref{g1int}), the SA leads to $\ln
(Q^2/\Lambda_{QCD}^2)$.

\section{Theoretical grounds for the shift $Q^2 \to Q^2+\mu^2$ in Eqs.~(\ref{gnsint})
 and (\ref{g1int})}

Both the singlet and non-singlet components of $g_1$ obey the
following Bethe-Salpeter equation depicted in Fig.~1:

\begin{equation}\label{bs}
g_1 = g_1^{Born}  + \imath \int \frac{d^4 k}{(2\pi)^4} (-2 \pi
\imath)\delta((q+k)^2 - m^2_q) \frac{2k_{\perp}^2}{(k^2 -
m^2_q)^2} M(p,k)
\end{equation}
where the $\delta$ -function (together with the factor $-2\pi
\imath$) corresponds to the cut propagator of the uppermost quark
with momentum $k$ and mass $m_q$ coupled to the virtual photon
lines and $2k_{\perp}^2$ appear after simplifying the spin
structure of the equation. $M(p,k)$ stands for the cut parton
ladders and the initial parton densities. In other words, $M(p,k)$
incorporates both the initial parton densities and radiative
corrections to them. This object can be called the polarized
parton distribution function. In the present paper we will address
it simply as the (parton) distribution, skipping other words. For
the sake of simplicity we dropped unessential numerical factors
($e^2_q/2$ for $f^{NS}$ and $<e^2_q>/2$ for the singlet) in
Eq.~(\ref{bs}). Obviously, $M$ in Eq.~(\ref{bs}) cannot depend on
$Q^2$.
\subsection{Prescription for the IR regularization of $M$}

The IR-divergent contributions to $M$ should  be regulated. We
follow the standard prescription and assign a fictitious mass
$\mu$ to the gluons in the IR-divergent propagators. In
particular, in the ladder graphs such propagators  are the
vertical ones. We assume that the value of $\mu$ satisfies
Eq.~(\ref{mu}). In contrast to the gluon ladders, quark ladders
are IR-stable because the quark mass $m_q$ acts as an IR cut-off.
In order to use the same cut-off $\mu$ for regulating both gluon
and quark IR divergences, we assume that, in addition to
Eq.~(\ref{mu}), $\mu \gg m_q$. After that $m_q$ can be dropped.
Therefore in order to regulate IR singularities, we should insert
$\mu^2$ in the IR -divergent, strut (vertical) propagators, for
both quarks and gluons. The horizontal propagators (rungs) are
IR-stable. This converts Eq.~(\ref{bs}) into

\begin{equation}\label{bsmu}
g_1 = g_1^{Born}  + \imath \int \frac{d^4 k}{(2\pi)^4} (-2 \pi
\imath)\delta((q+k)^2) \frac{2k_{\perp}^2}{(k^2 - \mu^2)^2} M(pk,
 k^2+\mu^2)
\end{equation}

\subsection{Solving the Bethe-Salpeter equation (\ref{bsmu})}

As the kinematics $x \ll 1$ is of the Regge type, we need to
express $M$ in the Regge kinematics as well. Let us
notice\footnote{We will consider such distributions in more detail
in next our paper.} that the expressions for $M$ can be obtained
from our results for $g_1$ in Ref.~\cite{egts,egtns} with
replacing the external photon virtuality $q^2 = -Q^2$ by the
external quark virtuality $k^2$ and $x$ by $-k^2/w\alpha$. In the
first place, we focus on applying Eq.~(\ref{bs}) to $g_1^{NS}$,
the non-singlet part of $g_1$ and denote $M^{NS}$ the involved
quark distribution. The expression for $M^{NS}$ accounting for the
total resummation of leading logarithmic contributions can also be
borrowed from our formula for $g_1^{NS}$ obtained in
Ref.~\cite{egtns}. This expression reads:
\begin{equation}\label{mns}
M^{NS}(p,k) = \int_{-\imath \infty}^{\imath \infty} \frac{d
\omega}{2 \pi \imath} \Big(\frac{2pk}{- k^2+\mu^2} \Big)^{\omega}
\omega f(\omega)\delta q(\omega)\Big(\frac{-
k^2+\mu^2}{\mu^2}\Big)^{H_{NS}(\omega)}~
\end{equation}
where $f  = 8 \pi^2 H_{NS},~~\delta q(\omega)$ is the initial
quark density in the $\omega$ -space and $H_{NS}$ is given by
Eq.~(\ref{hns}). $H_{NS}$ and $f$ include the total resummation of
leading logarithmic contributions. Similarly, the expression for
the singlet distribution $M^S$ can be obtained from our results
for $g_1^S$ in Ref.~\cite{egts}. It is convenient to rewrite
Eq.~(\ref{bsmu}) in terms of the Sudakov variables for momentum
$k$:

\begin{equation}\label{sud}
k = -\alpha q' + \beta p + k_{\perp}~,
\end{equation}
with $q' = q +xp,$ so that $q'^2 \approx p^2 \approx 0$.
Substituting Eq.~(\ref{mns}) into Eq.~(\ref{bsmu}), using the
Sudakov variables and changing the  order of the integrations,  we
obtain the following  equation for $g_1^{NS}$:

\begin{eqnarray}\label{bsns}
g_1^{NS} = g_1^{Born}+ \frac{1}{8\pi^2} \int_{-\imath
\infty}^{\imath \infty} \frac{d \omega}{2 \pi \imath}\omega
f(\omega)\delta q(\omega) \int d\alpha d\beta
 d k^2_{\perp}\frac{k^2_{\perp}}{(w\alpha\beta + k^2_{\perp} +\mu^2)^2}
 \delta(w\beta +  wx\alpha - w\alpha\beta -k^2_{\perp} -Q^2) \\
 \nonumber
 \Big(\frac{w\alpha}{w\alpha\beta
 +k^2_{\perp}+\mu^2}\Big)^{\omega}\Big(\frac{w\alpha \beta
+k^2_{\perp}+\mu^2}{\mu^2}\Big)^{H_{NS}}
\end{eqnarray}
where we have denoted $w  = 2pq$. As we consider $x \ll 1$, we can
neglect $x\alpha$ compared to $\beta$. Using the $\delta$
-function for integration over $\beta$, we arrive at
\begin{equation}\label{bsnsq}
g_1^{NS} = g_1^{Born}+ \frac{1}{8\pi^2} \int_{-\imath
\infty}^{\imath \infty} \frac{d \omega}{2 \pi \imath}\omega
f(\omega) \delta q(\omega) \int \frac{d \alpha d
k^2_{\perp}}{\alpha Q^2 + k^2_{\perp} + \mu^2}
\Big(\frac{w\alpha}{\alpha Q^2 +k^2_{\perp}+\mu^2}
 \Big)^{\omega}
 \Big(\frac{\alpha Q^2 +k^2_{\perp}+\mu^2}{\mu^2}\Big)^{H_{NS}}~.
\end{equation}
The integration region in  Eq.~(\ref{bsmu}) is shown in Fig.~2. It
is outlined by the following restrictions:
\begin{equation}\label{region}
w \gg k^2_{\perp} > 0,~w \gg w\alpha \gg \alpha Q^2
+k^2_{\perp}+\mu^2.
\end{equation}
Integrating over $\alpha$ and $k^2_{\perp}$ yields different
contributions, depending on the ratio between $\alpha Q^2$ and
$k^2_{\perp}$. The most important contribution comes from the
region $\emph{D}$ in Fig.~2. After integration  over $\alpha$ in
this region we get
\begin{eqnarray}\label{bsnsk}
g_1^{NS} = g_1^{Born}+ \frac{1}{8\pi^2} \int_{-\imath
\infty}^{\imath \infty} \frac{d \omega}{2 \pi \imath}\omega
f(\omega)\delta q(\omega) \frac{1}{\omega}\int_{Q^2}^{w} \frac{ d
k^2_{\perp}}{ k^2_{\perp}+\mu^2} \Big(\frac{w}{k^2_{\perp}+\mu^2}
 \Big)^{\omega} \Big(\frac{k^2_{\perp}+\mu^2}{\mu^2}\Big)^{H_{NS}}
 \\ \nonumber
=g_1^{Born}+ \frac{1}{8\pi^2} \int_{-\imath \infty}^{\imath
\infty} \frac{d \omega}{2 \pi \imath} f(\omega) \delta q(\omega)
 \int_{Q^2+\mu^2}^{w+\mu^2}
 \frac{ d t}{t} \Big(\frac{w}{t}
 \Big)^{\omega} \Big(\frac{t}{\mu^2}\Big)^{H_{NS}}
 ~.
\end{eqnarray}
The leading contribution in Eq.~(\ref{bsnsk}) comes from the
lowest limit $t = Q^2+ \mu^2$ and gives
\begin{equation}\label{bssol}
g_1^{NS} = g_1^{Born}+ \frac{1}{8\pi^2} \int_{-\imath
\infty}^{\imath \infty} \frac{d \omega}{2 \pi \imath}\frac{
f(\omega)\delta q(\omega)}{(\omega
-H_{NS})}\Big(\frac{w}{(Q^2+\mu^2)}
 \Big)^{\omega} \Big(\frac{Q^2+\mu^2}{\mu^2}\Big)^{H_{NS}}
\end{equation}
which proves the validity of the shift $Q^2 \to Q^2+\mu^2$
suggested in Ref.~\cite{egtsmq}. Let us brush up
Eq.~(\ref{bssol}). Replacing $f(\omega)$ by $8\pi^2H_{NS}$ we see
that in Eq.~(\ref{bssol})
\begin{equation}\label{hx}
\frac{H_{NS}}{\omega - H_{NS}}  = -1  + \frac{\omega}{\omega -
H_{NS}}~.
\end{equation}
The first term in Eq.~(\ref{hx}) cancels the Born contribution
$g_1^{Born}$ and the second term is the non-singlet coefficient
function (see Eq.~(\ref{hns})). Therefore, we arrive at
Eq.~(\ref{gnsint}) for the non-singlet $g_1^{NS}$ at small  $x$
and arbitrary $Q^2$. Eq.~(\ref{g1int}) for the singlet $g_1$ in
the same kinematic can be proved similarly.

\section{infrared power corrections at small $x$}

\subsection{Power corrections at large $Q^2$}

In the kinematics where $Q^2 > \mu^{2}$ and therefore $x
> z$, the terms with $Q^2 + \mu^2$ in Eqs.~(\ref{gnsint},\ref{g1int})
can be expanded into series in $\mu^2/Q^2$:
\begin{eqnarray}\label{qlarge}
\Big(\frac{1}{x + z}\Big)^{\omega}
\Big(\frac{Q^2 +\mu^2}{\mu^2}\Big)^{H_{NS}} &=&
\Big(\frac{1}{x}\Big)^{\omega}
\Big(\frac{Q^2}{\mu^2}\Big)^{H_{NS}} \Big[1 + \sum_{k = 1}
T_k^{NS}(\omega)\Big(\frac{\mu^2}{Q^2}\Big)^k\Big]~,\\ \nonumber
\Big(\frac{1}{x + z}\Big)^{\omega}
\Big(\frac{Q^2 +\mu^2}{\mu^2}\Big)^{\Omega_{\pm}} &=&
\Big(\frac{1}{x}\Big)^{\omega}\Big(\frac{Q^2}{\mu^2}\Big)^{\Omega_{\pm}}
\Big[ 1 + \sum_{k = 1}
T_k^{(\pm)}(\omega)\Big(\frac{\mu^2}{Q^2}\Big)^k\Big]
\end{eqnarray}
where
\begin{eqnarray}\label{tns}
T_k^{NS} &=& \frac{(-\omega + H_{NS})(-\omega + H_{NS} -
1)..(-\omega + H_{NS} - k + 1)}{k!},  \\ \nonumber
T_k^{(\pm)} &=& \frac{(-\omega + \Omega_{\pm})(-\omega + \Omega_{\pm}
-1)..(-\omega + \Omega_{\pm} -k + 1)}{k!}~.
\end{eqnarray}
It allows to rewrite Eqs.~(\ref{gnsint},\ref{g1int}) as follows:
\begin{eqnarray}\label{expqlarge}
g_1^{NS}(x+z,Q^2) &=& \widetilde{g}_1^{NS}(x,Q^2) +
\widetilde{g}_1^{NS}(x/y,Q^2) \otimes \sum_{k=1}
 ( \mu^2/Q^2)^k ~E_k^{NS}(y),\\ \nonumber
g_1^S(x+z,Q^2) &=&  \widetilde{g}_1^S(x,Q^2) + \sum_{k=1}
 (\mu^2/Q^2)^k~ \Big[\widetilde{g}_1^{(+)}(x/y,Q^2)
\otimes E_k^{(+)}(y) + \widetilde{g}_1^{(-)}(x/y,Q^2) \otimes
E_k^{(-)}(y)\Big]
\end{eqnarray}
where, using the conventional terms, $\widetilde{g}_1^{NS}$ and
$\widetilde{g}_1^S$ can be named the  non-singlet and singlet
components of the lowest twist contribution to $g_1$:
\begin{eqnarray}\label{gtilde}
\widetilde{g}_1^{NS} &=& (e^2_q/2) \int_{-\imath \infty}^{~\imath
\infty} \frac{d \omega}{2\pi\imath }\Big( \frac{1}{x}
\Big)^{\omega} C_{NS}(\omega) \delta q(\omega)
\Big(\frac{Q^2}{\mu^2} \Big)^{H_{NS}(\omega)}~~, \\ \nonumber
\widetilde{g}_1^{~S} &=& \widetilde{g}_1^{(+)} +
\widetilde{g}_1^{(-)} = \frac{<e^2_q>}{2} \int_{- \imath
\infty}^{~\imath \infty} \frac{d \omega}{2 \pi \imath}
\Big(\frac{1}{x} \Big)^{\omega}
 \Big[ \Big(C_q^{(+)}\Big( \frac{Q^2}{\mu^2}\Big)^{\Omega_{(+)}} +
C_q^{(-)} \Big( \frac{Q^2}{\mu^2}\Big)^{\Omega_{(-)}} \Big) \delta
q -  \\ \nonumber
& & \frac{A'}{2 \pi \omega^2} \Big(C_g^{(+)}
\Big( \frac{Q^2}{\mu^2}\Big)^{\Omega_{(+)} } + C_g^{(-)} \Big(
\frac{Q^2}{\mu^2}\Big)^{\Omega_{(-)}} \Big) \delta g \Big]
\end{eqnarray}
and
\begin{equation}\label{e}
E_k^{NS}(x) = \int_{- \imath \infty}^{\imath \infty} \frac{d
\omega}{2 \pi \imath} \Big( \frac{1}{x}\Big)^{\omega}
T_k^{NS}(\omega)~, \qquad E_k^{\pm}(x) = \int_{- \imath
\infty}^{\imath \infty} \frac{d \omega}{2 \pi \imath} \Big(
\frac{1}{x
}\Big)^{\omega} T_k^{\pm}(\omega)~.
\end{equation}


 The right-hand sides in
 Eq.~(\ref{expqlarge}) are the products of the power corrections
and $\widetilde{g}_1$. We call these corrections the infrared
power corrections. The functions
$\widetilde{g}_1^S,~\widetilde{g}_1^{NS}$ in Eq.~(\ref{gtilde})
were obtained in Refs.~\cite{egts,egtns} and they correspond to
the lowest twist contribution. They differ from the lowest twist
DGLAP expressions for $g_1$ by the total resummation of the
leading logarithms of $x$ and by the new parametrization of
$\alpha_s$ given by Eqss.~(\ref{a},\ref{aprime}). They include the
most important at small $x$ contributions of the LO and NLO DGLAP
formulas. When the lowest twist expressions of  Eq.~(\ref{gtilde})
are used for analysis of experimental data of the polarized DIS,
the power series in the rhs of Eq.~(\ref{expqlarge}) look as new
independent contributions. However, the left-hand sides of
Eq.~(\ref{expqlarge}) account for the total resummation of these
corrections. Finally, let us notice that the infrared power
corrections in Eq.~(\ref{expqlarge}) have nothing to do with the
standard parametrization $\alpha_s = \alpha_s(Q^2)$ as we do not
use it.

\subsection{Power corrections at small $Q^2$}

When $Q^2 < \mu^2$, $g_1^{NS}$ and $g_1^S$ cannot be expanded
similarly to Eq.~(\ref{expqlarge}). The power corrections for
small $Q^2$ are different. Indeed. in this case
\begin{eqnarray}\label{qsmall}
\Big(\frac{1}{x + z}\Big)^{\omega}\Big(\frac{Q^2 +
\mu^2}{\mu^2}\Big)^{H_{NS}} = \Big(\frac{1}{z}\Big)^{\omega}
 \Big[1 + \sum_{k = 1}
T_k^{NS}(\omega)\Big(\frac{Q^2}{\mu^2}\Big)^k\Big],
\\ \nonumber \Big(\frac{1}{x + z}\Big)^{\omega} \Big(\frac{Q^2 +
\mu^2}{\mu^2}\Big)^{\Omega_{\pm}} = \Big(\frac{1}{z}\Big)^{\omega}
\Big[ 1 + \sum_{k = 1}
T_k^{(\pm)}(\omega)\Big(\frac{Q^2}{\mu^2}\Big)^k\Big]~.
\end{eqnarray}
It leads to the following expressions for $g_1$ at small $Q^2$:
\begin{eqnarray}\label{expqsmall}
g_1^{NS}(x+z,Q^2) &=& \widetilde{g}_1^{NS}(z,\mu^2) +
\widetilde{g}_1^{NS}(z/y,\mu^2) \otimes \sum_{k=1}
 ( Q^2/\mu^2)^k~E_k^{NS}(y),
\\ \nonumber g_1^{~S}(x+z,Q^2) &=& \widetilde{g}_1^{S}(z,\mu^2) +
\sum_{k=1}
 (Q^2/\mu^2)^k~\Big[\widetilde{g}_1^{(+)}(z/y,\mu^2)
\otimes E_k^{(+)}(y) + \widetilde{g}_1^{(-)}(z/y,\mu^2) \otimes
E_k^{(-)}(y)\Big]
\end{eqnarray}
where the lowest twist contributions
$\widetilde{g}_1^{NS},~\widetilde{g}_1^{S}$ are given by
Eq.~(\ref{gtilde}) and $E_k^{NS},~E_k^{S}$ are defined in
Eq.~(\ref{e}). Both $\widetilde{g}_1^{NS}$ and
$\widetilde{g}_1^{S}$ do not depend on $x$ and $Q^2$. Instead,
they depend on the total energy $(p + q)^2$ of the process and are
constants when the $2pq$ is fixed. Both the $x$ and $Q^2$
-dependence are now associated with the power corrections.
Eqs.~(\ref{expqlarge},\ref{expqsmall}) show that the infrared
$Q^2$- corrections are different for large and small $Q^2$, so
that the series of Eq.~(\ref{expqsmall}) cannot be extrapolated
into the region of small $Q^2$ and similarly the series in
Eq.~(\ref{expqlarge}) cannot be extrapolated into the large $Q^2$-
region. Besides the terms with $(Q^2)^k$  given by
Eq.~(\ref{expqsmall}), similar contributions  can come from other
sources which are beyond our control. However, the $(Q^2)^k$
-terms in Eq.~(\ref{expqsmall}) are multiplied by the functions
$\widetilde{g}_1^{NS}$ and $\widetilde{g}_1^S$ which include the
total resummation of the leading logarithms and therefore they are
supposed to dominate, at small $x$, over the coefficients at the
other $(Q^2)^k$ -terms.

Although the large-$Q^2$ expansion (\ref{qlarge}) and the
small-$Q^2$ expansion (\ref{qsmall}) look quite similar, the power
corrections to $g_1^{NS}$ are actually  different for large and
small $Q^2$. It is easy to check that the linear in $\mu^2/Q^2$
term is present in the large-$Q^2$ expansion of
Eq.~(\ref{expqlarge}) for $g_1^{NS}$  while the term with
$Q^2/\mu^2$ is absent in the $g_1^{NS}$-expansion of
Eq.~(\ref{expqsmall}).

\section{Discussion}
At the region of small $x$, the DGLAP ordering (\ref{order})
should be lifted for accounting for leading logarithms of $1/x$,
so the infrared cut-off $\mu$ should be introduced explicitly to
regulate the infrared divergencies in every rung of the Feynman
graphs contributing to $g_1$. Similarly to DGLAP, $\mu$ can also
play the role of the starting point of the $Q^2$ -evolution,
though not obligatory. With both the perturbative and
non-perturbative contributions accounted for, $g_1$ does not
depend on $\mu$. However, this dependence does exist when the
non-perturbative contributions are neglected or accounted for
implicitly through the fits for the initial parton densities. In
this case the structure function $g_1$ depends on the value of
$\mu$ and the way it has been introduced. Introducing $\mu$ as the
fictitious mass inserted into the IR -divergent propagators, leads
to Eqs.~(\ref{gnsint},\ref{g1int}) suggested in Ref.~\cite{egtsmq}
for $g_1$. The expressions of Eqs.~(\ref{gnsint},\ref{g1int})
include the total resummation of double-logarithms and the most
important part of single-logarithms of $x$. They are obtained from
our previous results with  the shift $Q^2 \to Q^2+\mu^2$. The
theoretical grounds for such a shift are given by
Eqs.~(\ref{b}-\ref{bssol}).
Eqs.~(\ref{gnsint},\ref{g1int}), in contrast to DGLAP, can be used
both for large and small $Q^2$. Having been expanded into the
series in $1/(Q^2)^k$ at large $Q^2$ (or into series in $Q^2$ at
small $Q^2$), Eqs.~(\ref{gnsint},\ref{g1int}) yield the power
$Q^2$- corrections.  The series of the corrections are represented
by expressions (\ref{expqlarge},\ref{expqsmall}). The power series
of Eqs.~(\ref{expqlarge},\ref{expqsmall}) for large and small
$Q^2$ are derived from the same formulas. However after the
expansion has been made, they cannot be related to each other with
simply varying $Q^2$.  We suggest that accounting for the new
source of the power contributions given by
Eqs.~(\ref{gnsint},\ref{g1int}) can sizably change the
conventional analysis of the higher twists contributions to the
Polarized DIS because such contributions appear in the present
analysis of experimental data as a discrepancy between the
experimental data and the Standard Approach predictions. Let us
remind that in Ref.~\cite{egtinp} we showed that the singular
$(\sim x^{- \alpha})$ factors in the standard fits mimic the total
resummations of $\ln^k x$, i.e. they have a purely perturbative
origin contrary to previous common expectations.  Similarly, a
good portion of the commonly believed non-perturbative power
corrections in the conventional analysis of experimental data
 can actually be of the perturbative infrared origin. However being
misinterpreted as non-perturbative terms, they can mimic the power
expansion in Eq.~(\ref{expqlarge}). In particular,
Eq.~(\ref{gnsint}) predicts that the power $\sim 1(Q^2)^k$
--corrections to $g_1^{NS}$ should appear at $Q^2 \gtrsim
1$~GeV$^2$ and cannot appear at smaller values of $Q^2$. It agrees
with the phenomenological  observations obtained in
Refs.~\cite{sid} from conventional analysis of experimental data.
On the other hand, Eq.~(\ref{g1int}) predicts that the similar
power corrections
 to the singlet $g_1$ should be seen at greater values of
$Q^2$. Clearly, the use of Eqs.~(\ref{gnsint},\ref{g1int}) for the
lower twist contributions to $g_1$, instead of DGLAP, would allow
one to revise the impact of the genuine higher twists
contributions which are known to be of the non-perturbative
origin.

Finally, we would like to remind that our results explicitly
depend on the infrared cut-off $\mu$. As pointed out in Sect.~II,
such a dependence would vanish if analytic expressions for the
probabilities $\Phi_{q,g}$ were obtained and used in
Eqs.~(\ref{gnsint},\ref{g1int}) instead of $\delta q$ and $\delta
g$.

\section{Acknowledgements}
We are grateful to R.~Windmolders and A.~Korzenev for drawing our
attention to the problem of the power corrections to $g_1$ at
small $x$. We are also  indebted to G.P.~Korchemsky and
S.I.~Alekhin for discussing the power corrections to $g_1$ in the
framework of the Standard Approach. The work is supported in part
by the Russian State Grant for Scientific School
RSGSS-5788.2006.2.

\begin{figure}
\begin{center}
\begin{picture}(340,200)
\put(0,0){
\epsfbox{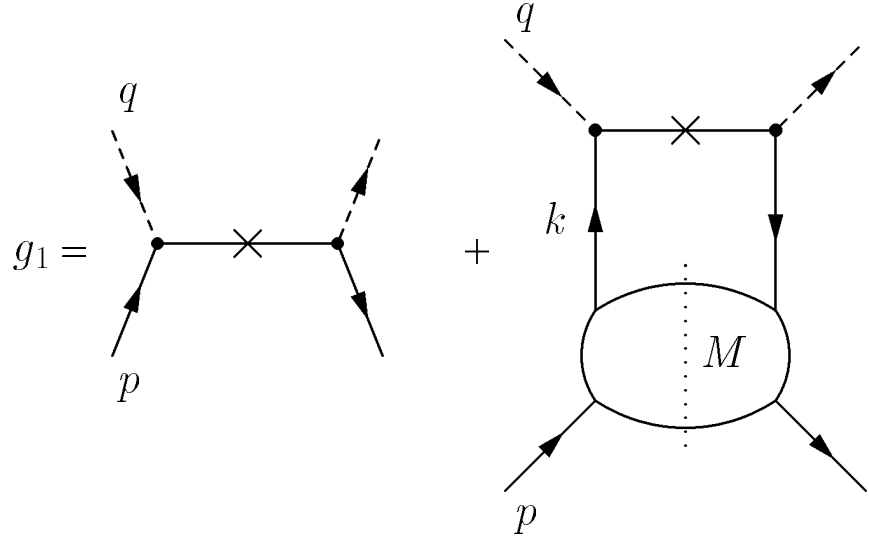} }
\end{picture}
\end{center}
\caption{The Bethe-Salpeter equation for $g_1$.}
\end{figure}
\begin{figure}
\begin{center}
\begin{picture}(250,170)
\put(0,0){
\epsfbox{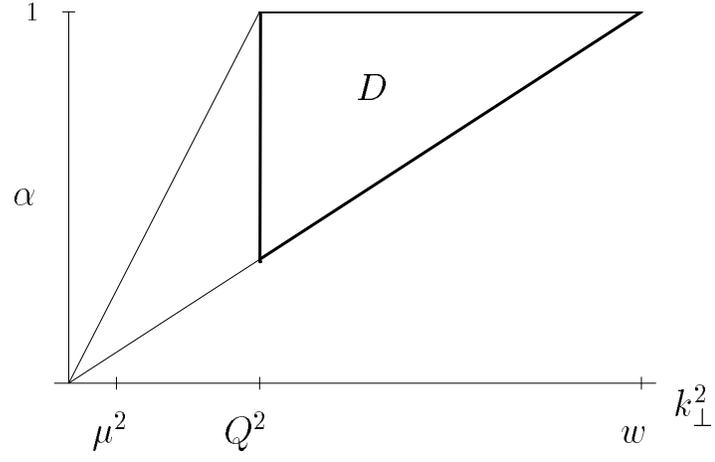} }
\end{picture}
\end{center}
\caption{The integration region over $\alpha$ and $k^2_{\perp}$ in
Eq.~(\ref{bsnsq}).}
\end{figure}

\end{document}